# Spin Waves in Thin Films and Magnonic Crystals with Dzyaloshinskii–Moriya Interactions

**Rodolfo A. Gallardo,[a,b] David Cortés-Ortuño,[c] Roberto E. Troncoso,[d] and Pedro Landeros[a,b]**

[a]*Departamento de Física, Universidad Técnica Federico Santa María, Avenida España 1680, Valparaíso, Chile*
[b]*Center for the Development of Nanoscience and Nanotechnology (CEDENNA), 917-0124 Santiago, Chile*
[c]*Faculty of Engineering and Physical Sciences, University of Southampton, Southampton SO17 1BJ, United Kingdom*
[d]*Center for Quantum Spintronics, Department of Physics, Norwegian University of Science and Technology, NO-7491 Trondheim, Norway*
pedro.landeros@usm.cl

## 5.1 Introduction

In the late fifties of the past century, Dzyaloshinskii proposed in a seminal paper a phenomenological theory of antisymmetric exchange coupling between spins to explain the phenomenon of weak ferromagnetism in antiferromagnetic compounds [1]. Two years later, Moriya derived this interaction as a spin–orbit

**2** *Spin Waves in Thin Films and Magnonic Crystals with Dzyaloshinskii–Moriya Interactions*

coupling between electrons within the framework of superexchange theory [2–4]. It was then shown that this anisotropic exchange interaction arises in materials that lack inversion symmetry and where strong spin–orbit coupling effects are present. Nowadays, this antisymmetric exchange coupling is known as the Dzyaloshinskii–Moriya interaction (DMI) and has been a key ingredient for the description of the magnetic properties of a variety of compounds with broken symmetry [5–10]. These include noncentrosymmetric bulk ferromagnets, multiferroics, perovskites, cuprates, and ferromagnetic thin films [11–16], among others. The study of materials with DMIs has been pursued with high interest because it has been thoroughly established, both by theory and experiment, that DMIs induce chiral, topological, and nonreciprocal features. A primary consequence is the occurrence of chiral spin textures such as magnetic helices, skyrmions, skyrmion lattices, and chiral domain walls (DWs) in ferromagnetic materials [17–42].

    Several theories have been developed to understand the origin of interfacial DMI in ultrathin ferromagnetic films in contact with a heavy metal. Originally, Levy and Fert [43–46] developed a theory for disordered magnetic alloys with heavy metal impurities that involves an additional contribution to the Ruderman–Kittel–Kasuya–Yosida (RKKY) interaction [47, 48]. This extra term is of DMI type and arises from the spin–orbit scattering of the conduction electron gas with the heavy metal impurities [45]. Recently, it has been proposed that the origin of the interfacial DMI is closely related to the proximity-induced magnetic moment in the heavy metal [49], while ab initio calculations reveal that the proximity-induced magnetic moment in the heavy metal has no direct correlation with the DMI in Co/Pt interfaces [50]. Using a density functional approach, Belabbes et al. [51] demonstrated that the sign and magnitude of the DMI are related to the degree of 3d–5d orbital hybridization around the Fermi level. The temperature dependence of the DMI and its interplay with the anisotropy of the orbital magnetic moment and the magnetic dipole moment in Pt/Co/MgO trilayers were studied experimentally and theoretically by Kim et al. [52], where the interfacial DMI originates from the asymmetric charge distribution caused by the breaking of inversion symmetry. According to these studies, the underlying physical mechanism

behind the DMI in ferromagnetic ultrathin films is not completely understood and is still a topic under discussion.

Among the different kinds of elementary excitations in condensed matter physics, spin waves play a significant role because of the shorter wavelengths that can be attained at GHz frequencies and also due to the reduced losses by heating. Spin waves have been observed from millimeter down to micrometer [53] and, more recently, to submicrometer length scales [54–62]. These spin excitations can be generated using microwaves in resonant cavities, alternating Oersted fields produced by radiofrequency currents, inelastic neutron scattering, spin-transfer torques from electric currents, visible light as in Brillouin light scattering (BLS), and the time-resolved magneto-optical Kerr effect [63–67]. A more sophisticated experimental prove of spin waves requires a synchrotron and is based on time-resolved scanning transmission X-ray microscopy [60–62].

The influence of the DMI on the spin-wave spectra has been widely studied in the last years. The first observation of nonreciprocal spin waves [16, 68, 69] caused by interfacial DMI was achieved by Zakeri et al. [70, 71] using the spin-polarized electron energy loss (SPEEL) experimental technique. A few years later, nonreciprocal spin waves were observed by means of BLS spectroscopy in several ferromagnet/heavy metal interfaces [72–87]. In agreement with the results of Zakeri et al. [70, 71], these numerous experiments have shown that interfacial DMI creates a noticeable frequency asymmetry in the spin-wave dispersion of counterpropagating Damon–Eshbach (DE) spin waves. The DMI-induced magnon asymmetry has been theoretically established by several groups [88–101]. From these calculations, one can see that the strength of the DMI can be estimated experimentally by measuring the frequency difference of counterpropagating waves, as proposed in Ref. [96]. Furthermore, it was also predicted that in systems with bulk DMI a frequency difference could be observed in the backward-volume geometry [91, 96], that is, with the magnetization and wave vector parallel to each other and along the plane of the film. This effect has been measured in a few chiral lattice magnets [69, 102–105]. Beyond the study of spin waves in field-polarized states, spin excitations in non-uniform magnetic textures, such as helical, conical, and skyrmion phases, have also been reported [106–111].

The dynamics of spin waves under the influence of bulk and interfacial DMI, both in thin films and magnonic crystals (MCs), is reviewed in this chapter. Nonreciprocal properties induced by different classes of DMI in thin-film magnets are discussed in Sections 5.2 and 5.3. A detailed analysis from the point of view of micromagnetic simulations is given in Section 5.4. The band structure of a chiral MC defined with a periodic DMI is discussed in Section 5.5, where both the plane-wave method and micromagnetic simulations are used to theoretically analyze the system. Conclusions are summarized in Section 5.6.

## 5.2 Spin Waves in Thin Films with Bulk DMI

In this section, a theoretical background is given for the spin-wave dynamics in thin films with bulk DMI. It is known that two counterpropagating spin waves show a frequency difference in magnetic films with bulk DMI but in the backward-volume geometry [91, 96], that is, with the magnetization and wave vector both parallel to each other and along the film's plane. This behavior has been measured in some chiral lattice magnets [102–105, 112] and also in the noncentrosymmetric antiferromagnet $\alpha\text{Cu}_2\text{V}_2\text{O}_7$ [113].

The dynamics of the magnetization is described theoretically by following Refs. [96, 114, 115], where the magnetization is composed by a static and a dynamic part: $\mathbf{M}(\mathbf{r}; t) = M_Z \widehat{Z} + \mathbf{m}(\mathbf{r}; t)$. The static part points always along the $Z$ axis, which is chosen along the equilibrium magnetization orientation (see Fig. 5.1). The dynamic magnetization can be written as $\mathbf{m}(\mathbf{r}; t) = \mathbf{m}(\mathbf{r})e^{i\omega t}$ so that the spatial term is $\mathbf{m}(\mathbf{r}) = m_X(\mathbf{r})\widehat{X} + m_Y(\mathbf{r})\widehat{Y}$. For small deviations of the magnetization from the equilibrium position, it can be assumed that the dynamic components satisfy $m_{X,Y} \ll M_s$ and then $M_Z \approx M_s - (m_X^2 + m_Y^2)/(2M_s)$ [114]. The dynamic magnetization components are represented in Fourier form in terms of the wave vector $\mathbf{k}$ as

$$m_{\alpha,\beta}(\mathbf{r}) = \frac{1}{\sqrt{L^2 d}} \sum_{\mathbf{k}} m_{\alpha,\beta}(\mathbf{k}) e^{i\mathbf{k}\cdot\mathbf{r}}, \tag{5.1}$$

where **r** and **k** lie in the plane of the ferromagnetic film of area $L^2$ and thickness $d$, and $\alpha, \beta = X, Y$. The spin wave Hamiltonian can be expressed as [114, 115]

$$\mathcal{H} = \frac{1}{2M_s} \sum_{\mathbf{k}} \sum_{\alpha,\beta} \mathcal{W}_{\alpha,\beta}(\mathbf{k}) m_\alpha^*(\mathbf{k}) m_\beta(\mathbf{k}), \quad (5.2)$$

where $\mathcal{W}_{\alpha,\beta}(\mathbf{k})$ are the stiffness fields, which contain contributions from the dipolar energy, Zeeman energy, exchange interaction, anisotropies, and the DMI. These energies were calculated in Refs. [114, 115], while the role of DMI was studied in Refs. [96, 110] for several Lifshitz invariants representing different crystal symmetries [12, 17–19, 116].

In the micromagnetic limit, the DMI Hamiltonian can be written as

$$\mathcal{H}_{\text{DM}}(\mathbf{r}) = \int_V w_{\text{DM}} dV, \quad (5.3)$$

where the energy density $w_{\text{DM}}$ involves combinations of first-order spatial derivatives of the magnetization known as *Lifshitz invariants* [12, 17–19], which are defined as

$$\mathcal{L}_{ij}^{(k)} = M_i \frac{\partial M_j}{\partial x_k} - M_j \frac{\partial M_i}{\partial x_k}. \quad (5.4)$$

Depending on the symmetry of the noncentrosymmetric crystal, $w_{\text{DM}}$ assumes different forms. One commonly studied sum of invariants has the structure (see page 18 in Ref. [116])

$$\mathcal{L} = \mathcal{L}_{yx}^{(z)} + \mathcal{L}_{xz}^{(y)} + \mathcal{L}_{zy}^{(x)} = \mathbf{M} \cdot (\nabla \times \mathbf{M}), \quad (5.5)$$

that is characteristic of noncentrosymmetric cubic crystals, for instance, MnSi, FeGe, MnGe, or $Cu_2OSeO_3$ [110, 116]. The DM energy density is given by $w_{\text{DM}} = \mu_0 \lambda_{\text{DM}} \mathcal{L}$, where $\lambda_{\text{DM}} = D/\mu_0 M_s^2$ has units of length, and is proportional to the usual DMI strength $D$ in units of Joule/meter$^2$. For cubic crystals, the DM energy reads

$$\mathcal{H}_{\text{DM}}(\mathbf{r}) = \frac{D}{M_s^2} \int_V \mathbf{M} \cdot (\nabla \times \mathbf{M}) dV. \quad (5.6)$$

For different crystal symmetries, one can write the DMI in terms of other combinations of Lifshitz invariants, as shown in Table 5.1, which was adapted from Ref. [96], where cgs units were used with the strength of the DMI represented by the length $\lambda_{\text{DM}}$. In Table 5.1, as in the rest of the chapter, SI units are used, while the strength of the DMI is represented by the parameter $D$ [97, 110]. Following Refs. [96, 115], the magnetization components in the fixed coordinate system $xyz$ (see Fig. 5.1) are specified as

$$M_x = m_X(\mathbf{r}) \tag{5.7a}$$
$$M_y = M_Z \sin\phi_{\mathrm{m}} + m_Y(\mathbf{r}) \cos\phi_{\mathrm{m}} \tag{5.7b}$$
$$M_z = M_Z \cos\phi_{\mathrm{m}} - m_Y(\mathbf{r}) \sin\phi_{\mathrm{m}}, \tag{5.7c}$$

where $\phi_{\mathrm{m}}$ is the angle between the equilibrium magnetization and the plane of the film.

**Table 5.1** DM energy density and corresponding asymmetry in the spin-wave dispersion relation for different symmetry classes. The case of interfacial DMI corresponds to the $C_{nv}$ symmetry group [96].

| Symmetry class | Energy density $w_{DM}$ | Frequency asymmetry $\Delta f$ |
|---|---|---|
| **T** | $(D/M_s^2)\left(\mathcal{L}_{yx}^{(z)} + \mathcal{L}_{xz}^{(y)} + \mathcal{L}_{zy}^{(x)}\right)$ | $2\gamma D (\pi M_s)^{-1} k \cos\phi_{\mathbf{k}} \cos\phi_{\mathrm{m}}$ |
| **$C_{nv}$** | $(D/M_s^2)\left(\mathcal{L}_{zy}^{(z)} + \mathcal{L}_{xy}^{(x)}\right)$ | $-2\gamma D (\pi M_s)^{-1} k \sin\phi_{\mathbf{k}} \cos\phi_{\mathrm{m}}$ |
| **$D_{2d}$** | $(D/M_s^2)\left(\mathcal{L}_{xy}^{(z)} + \mathcal{L}_{zy}^{(x)}\right)$ | $-2\gamma D (\pi M_s)^{-1} k \cos\phi_{\mathbf{k}} \cos\phi_{\mathrm{m}}$ |
| **$D_n$** | $(D^{(1)}/M_s^2)\left(\mathcal{L}_{zy}^{(x)} - \mathcal{L}_{xy}^{(z)}\right)$ $+ (D^{(2)}/M_s^2)\left(\mathcal{L}_{zx}^{(y)}\right)$ | $-2\gamma D^{(1)} (\pi M_s)^{-1} k \cos\phi_{\mathbf{k}} \cos\phi_{\mathrm{m}}$ |
| **$C_n$** | $(D^{(1)}/M_s^2)\left(\mathcal{L}_{zy}^{(z)} + \mathcal{L}_{xy}^{(x)}\right)$ $+ (D^{(2)}/M_s^2)\left(\mathcal{L}_{zy}^{(x)} - \mathcal{L}_{xy}^{(z)}\right)$ | $-2\gamma (\pi M_s)^{-1} k[D^{(1)} \sin\phi_{\mathbf{k}}$ $- D^{(2)} \cos\phi_{\mathbf{k}}]\cos\phi_{\mathrm{m}}$ |

By expanding the dynamic components of the magnetization up to second order, the sum of invariants (see Eq. 5.5) are expressed as

$$\mathcal{L} = M_s \frac{\partial m_Y}{\partial x} + M_s \sin\phi_{\mathrm{m}} \frac{\partial m_X}{\partial z} + m_Y \cos\phi_{\mathrm{m}} \frac{\partial m_X}{\partial z} - m_X \cos\phi_{\mathrm{m}} \frac{\partial m_Y}{\partial z}. \tag{5.8}$$

In the thin-film limit, the dynamic magnetization components do not depend on the normal coordinate $y$, so $m_{X,Y}(\mathbf{r}) = m_{X,Y}(x, z)$, and then it is possible to integrate along the thickness to obtain

$$\mathcal{H}_{\mathrm{DM}} = \frac{Dd \cos\phi_{\mathrm{m}}}{M_s^2} \int \left[ m_Y \frac{\partial m_X}{\partial z} - m_X \frac{\partial m_Y}{\partial z} \right] dx\, dz, \tag{5.9}$$

where the linear terms have been disregarded, since they cancel out due to the equilibrium condition. Using the Fourier representation

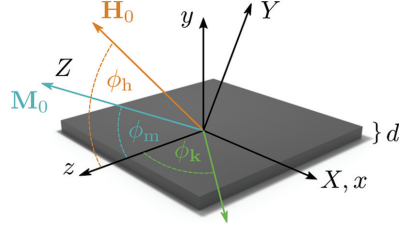

**Figure 5.1** (a) Geometry and notation. Vectors $\mathbf{M_0}$ and $\mathbf{H_0}$ correspond, respectively, to the equilibrium magnetization and applied field, while $d$ denotes the thickness of the ferromagnetic film. The coordinate system $(x, y, z)$ is fixed in the film, while the coordinates $(X, Y, Z)$ are defined according to the equilibrium magnetization, such that $Z$ is always parallel to $\mathbf{M_0}$, which makes an angle $\phi_m$ with the film. Spin waves propagate in the plane at an angle $\phi_\mathbf{k}$ with the $z$ axis.

(see Eq. 5.1) the Hamiltonian associated to the bulk DMI is given by

$$\mathcal{H}_{\text{DM}} = \frac{1}{2M_s} \sum_{\mathbf{k}} [\mathcal{W}_{XY}^{\text{DM}}(\mathbf{k})\, m_X^*(\mathbf{k})\, m_Y(\mathbf{k})$$
$$+ \mathcal{W}_{YX}^{\text{DM}}(\mathbf{k})\, m_Y^*(\mathbf{k}) m_X(\mathbf{k})], \quad (5.10)$$

in which the DMI stiffness fields are given by

$$\mathcal{W}_{XY}^{\text{DM}}(\mathbf{k}) = -\frac{2D}{M_s} ik\, \cos\phi_\mathbf{k}\, \cos\phi_m, \quad (5.11)$$

and $\mathcal{W}_{XY}^{\text{DM}}(\mathbf{k}) = -\mathcal{W}_{YX}^{\text{DM}}(\mathbf{k})$. Here, $k$ is the magnitude of the wave vector and $\phi_\mathbf{k}$ is the angle between the wave vector and the projection of the equilibrium magnetization onto the plane, as shown in Fig. 5.1. Using the other energetic contributions (see Refs. [96, 115] for details) that add to the total Hamiltonian (Eq. 5.2), which include the DM term (Eq. 5.10), the stiffness fields are calculated as

$$\mathcal{W}_{XX}(\mathbf{k}) = \mathcal{W}_{XX}(0) + \mu_0 M_s F(kd) \sin^2\phi_\mathbf{k} + D_{\text{ex}}k^2, \quad (5.12\text{a})$$

$$\mathcal{W}_{YY}(\mathbf{k}) = \mathcal{W}_{YY}(0) - \mu_0 M_s F(kd) [\cos 2\phi_m +$$
$$\sin^2\phi_\mathbf{k}\sin^2\phi_m] + D_{\text{ex}}k^2, \quad (5.12\text{b})$$

$$\mathcal{W}_{XY}(\mathbf{k}) = -\mu_0 M_s F(kd) \sin\phi_\mathbf{k} \cos\phi_\mathbf{k} \sin\phi_m$$
$$- \frac{2D}{M_s} ik \cos\phi_\mathbf{k} \cos\phi_m, \quad (5.12\text{c})$$

$$\mathcal{W}_{YX}(\mathbf{k}) = -\mu_0 M_s F(kd) \sin\phi_\mathbf{k} \cos\phi_\mathbf{k} \sin\phi_m$$
$$+ \frac{2D}{M_s} ik \cos\phi_\mathbf{k} \cos\phi_m, \quad (5.12\text{d})$$

where $D_{\text{ex}} = 2A/M_s$ is the exchange stiffness and $F(x) = 1 - (1 - e^{-x})/x$. Additionally,

$$\mathcal{W}_{XX}(0) = \mu_0 H_0 \cos(\phi_h - \phi_m) - \mu_0 M_{\text{eff}} \sin^2\phi_m, \quad (5.13a)$$
$$\mathcal{W}_{YY}(0) = \mu_0 H_0 \cos(\phi_h - \phi_m) + \mu_0 M_{\text{eff}} \cos 2\phi_m. \quad (5.13b)$$

In Eqs. 5.13, $H_0$ is a static magnetic field applied at an angle $\phi_h$ respect to the plane, and $M_{\text{eff}} = M_s - H_s$ is the effective magnetization, with $H_s$ the perpendicular surface anisotropy. The angle $\phi_m$ can be calculated from a simple equilibrium condition [96, 115].

To calculate the spin-wave dispersion relation, the dynamic magnetization components are promoted to operators, with well-known commutation rules [114, 117]. The temporal evolution of the dynamic magnetization is given by

$$i\hbar \dot{m}_{X,Y}(\mathbf{k};t) = [m_{X,Y}(\mathbf{k};t), \mathcal{H}]. \quad (5.14)$$

Then, using $m_{X,Y}(-\mathbf{k};t) = m^*_{X,Y}(\mathbf{k};t)$, the Hamiltonian given by Eqs. 5.2 and 5.12, the commutation rules from Refs. [114, 117] and the definition of the absolute value of the gyromagnetic ratio $\gamma = -\mu\hbar^{-1}$, with $\mu$ as the magnetic moment of a magnetic ion in the film, it is possible to obtain a pair of first-order differential equations that can be written in matrix form,

$$\begin{pmatrix} \dot{m}_X \\ \dot{m}_Y \end{pmatrix} = \gamma \begin{pmatrix} -\mathcal{W}_{YX} & -\mathcal{W}_{YY} \\ \mathcal{W}_{XX} & \mathcal{W}_{XY} \end{pmatrix} \begin{pmatrix} m_X \\ m_Y \end{pmatrix}. \quad (5.15)$$

The dispersion relation of the spin waves can now be computed by assuming harmonic solutions for the dynamic magnetization components and reads

$$\omega_{\pm}(\mathbf{k}) = \gamma \left\{ -\mathcal{I}[\mathcal{W}_{XY}] \pm \sqrt{\mathcal{W}_{XX}\mathcal{W}_{YY} - \mathcal{R}[\mathcal{W}_{XY}]^2} \right\}. \quad (5.16)$$

Here, $\mathcal{I}[\mathcal{W}_{XY}]$ and $\mathcal{R}[\mathcal{W}_{XY}]$ denote the imaginary and the real part of the function $\mathcal{W}_{XY}$. Notice that the above formula is valid for any symmetry class, provided that $\mathcal{W}_{XY}(\mathbf{k}) = \mathcal{W}^*_{YX}(\mathbf{k})$ [96]. For a crystal with symmetry class T, the dispersion relation becomes

$$\omega(\mathbf{k}) = \frac{2\gamma D}{\mathcal{M}_s} k \cos\phi_{\mathbf{k}} \cos\phi_m + \gamma \left[ \mathcal{W}_{XX}(\mathbf{k}) \mathcal{W}_{YY}(\mathbf{k}) \right.$$
$$\left. - \left( \frac{\mu_0 M_s}{2} kd \sin\phi_{\mathbf{k}} \cos\phi_{\mathbf{k}} \sin\phi_m \right)^2 \right]^{\frac{1}{2}}. \quad (5.17)$$

The DMI appears only in the linear term proportional to *D*. For the Lifshitz invariants describing chiral lattice ferromagnets (symmetry class T in Table 5.1), the influence of the DMI on the spectrum is strongest when the magnetization lies in the plane ($\phi_\mathrm{m} = 0$) and when wave vectors are parallel to the equilibrium magnetization ($\phi_\mathbf{k} = 0$), which are known as backward-volume magnetostatic spin waves. Under wave vector inversion $\mathbf{k} \to -\mathbf{k}$, where $\mathbf{k}$ rotates $\pi$ radians, one must change $\phi_\mathbf{k}$ by $\phi_\mathbf{k} \pm \pi$. In this case, the DM term in the dispersion relation changes its sign, since $\cos(\phi_\mathbf{k} \pm \pi) = -\cos\phi_\mathbf{k}$, and therefore, because of this asymmetry, the relation is nonreciprocal [91, 96, 102, 104, 110]. According to this, the frequency $f(\mathbf{k}) = \omega(\mathbf{k})/(2\pi)$ can be used to define a frequency asymmetry given by

$$\Delta f = f(\mathbf{k}) - f(-\mathbf{k}). \tag{5.18}$$

In the case of the symmetry class T, the nonreciprocity in the spin-wave dispersion relation is linear with respect to the wave vector and reads

$$\Delta f = \frac{2\gamma D}{\pi M_\mathrm{s}} k \cos\phi_\mathbf{k} \cos\phi_\mathrm{m}. \tag{5.19}$$

For this particular crystallographic class, and in the backward-volume configuration $\phi_\mathbf{k} = 0$, nonreciprocal spin waves have been observed recently in different magnetic compounds. These include the bulk noncentrosymmetric ferromagnets $LiFe_5O_8$ [102], $Cu_2OSeO_3$ [104], MnSi [103, 105, 112], and the noncentrosymmetric antiferromagnet $\alpha Cu_2V_2O_7$ [113].

Similar results are also obtained for the other allowed combinations of Lifshitz invariants, which are governed by the symmetry of the crystallographic classes. The $\Delta f$ frequency expressions of the different invariants are summarized in Table 5.1. They follow a common structure because in the thin-film limit, there is no dependence of the dynamic magnetization on the *y* coordinate and the second-order terms share the same structure. Furthermore, as Eq. 5.19 dictates, all the asymmetries vanish if the system is saturated perpendicularly to the film plane ($\phi_\mathrm{m} = \pi/2$), and $\Delta f$ is largest when **M** is parallel to the film plane. The differences in the frequency asymmetries between different symmetry classes are the dependence on the angle $\phi_\mathbf{k}$ between $\mathbf{M}$ and $\mathbf{k}$ and the type of DMI constants involved.

The full set of response functions for microwave excitations was also calculated (see Eqs. 34–37 in Ref. [96]), which allows us to estimate the nonreciprocal absorption spectrum in systems with bulk and interfacial DMIs.

## 5.3 Spin Waves in Thin Films with Interfacial DMI

A different mechanism for the presence of DMI in ferromagnets is the spin–orbit coupling arising at the interface of a thin-film ferromagnet adjacent to a heavy metal layer. Under the presence of interfacial DMI, spin waves exhibit similar effects than the ones in bulk DMI systems. These effects have been theoretically studied by several groups [86, 88–101]. Furthermore, in these film samples, of particular interest is the dependence of the interfacial DMI strength [86] on the heavy metal thickness, which requires the evaluation of the DMI vector. The thickness dependence of the DMI vector and the frequency nonreciprocity in interfacial DMI films are discussed in this section.

Starting with the analysis of the heavy metal thickness, the theoretical approach used here is based on the microscopic DMI Hamiltonian, which reads

$$\mathcal{H}_{\mathrm{DM}} = \sum_{\langle i, j \rangle} \mathbf{D}_{ij} \cdot \left( \mathbf{S}_i \times \mathbf{S}_j \right). \qquad (5.20)$$

The DMI couples any pair of neighboring atomic spins $\mathbf{S}_i$ and $\mathbf{S}_j$ in the interfacial layer of the ferromagnetic film through a third heavy metal site. The indexes $i$ and $j$ label a pair of interacting ferromagnetic spins at the interface. The DM vector $\mathbf{D}_{ij}$ links ferromagnetic spins at sites $i$ and $j$ with a third heavy metal site, labeled by a lattice vector $\mathbf{l}$ in the heavy metal, that is perpendicular to the triangle described by the three sites [11, 12]. This theory follows the formalism developed by Fert and Levy for disordered magnetic alloys with heavy metal impurities [45, 46]. The existence of a DMI at the ferromagnet/heavy metal interface is a well-established phenomenon, in which the DMI strength decreases with the thickness $d$ of the ferromagnetic layer, due to the interface nature of the DMI [75], and increases up to a saturation value with the thickness of the heavy metal layer [86, 87]. This behavior suggests that there is a significant number of heavy metal atoms contributing to the strength of the interfacial DMI. Therefore, to calculate the DM vector, the thickness and lattice structure of the heavy metal need to be considered. The DM vector that links a pair of ferromagnetic spins includes contributions from a group of neighboring heavy metal atoms and can be evaluated by summing over several heavy metal lattice vectors **l** (see Fig. 1 in Ref. [86]), that is,

$$\mathbf{D}_{ij} = \sum_{\mathbf{l}} V_1 \frac{\hat{\mathbf{R}}_i^{\mathbf{l}} \cdot \hat{\mathbf{R}}_j^{\mathbf{l}}}{R_i^{\mathbf{l}} R_j^{\mathbf{l}} R_{ij}} \hat{\mathbf{R}}_i^{\mathbf{l}} \times \hat{\mathbf{R}}_j^{\mathbf{l}} \sin\left[k_\mathrm{F}\left(R_i^{\mathbf{l}} + R_j^{\mathbf{l}} + R_{ij}\right) + \frac{\pi Z_d}{10}\right], \quad (5.21)$$

where $V_1 = [(135\pi \lambda_d \Gamma^2)/(32 E_\mathrm{F}^2 k_\mathrm{F}^3)] \sin(\pi Z_d/10)$ has units of Joule m$^3$ and depends on the spin–orbit coupling constant $\lambda_d$, the Fermi energy $E_\mathrm{F}$ and Fermi wave vector $\mathbf{k}_\mathrm{F}$ of the conduction electrons, the coupling constant $\Gamma$ between ferromagnetic spins, and the number of incomplete subshell electrons $Z_d$ (see Eqs. 5 and 6 in Ref. [45] and Eq. 26 in Ref. [46] for details). The position vectors $\mathbf{R}_i^{\mathbf{l}}$ and $\mathbf{R}_j^{\mathbf{l}}$, join the ferromagnetic sites $\mathbf{S}_i$ and $\mathbf{S}_j$ to the heavy metal sites. The DM vector between spins at sites $i$ and $i + x$ can be written as $\mathbf{D}_{i,j=i+x} = -D\hat{\mathbf{z}}$, while the corresponding DM vector between spins at $i$ and $i + z$ reads $\mathbf{D}_{i,j=i+z} = D\hat{\mathbf{x}}$ [86]. The $y$ component of $\mathbf{D}_{ij}$ vanishes due to the assumed perfectly flat interface.

The calculation of the spin-wave dispersion relation and the frequency nonreciprocity follows the same approach of the previous section but with a different combination of Lifshitz invariants, namely $\mathcal{L} = \mathcal{L}_{zy}^{(z)} + \mathcal{L}_{xy}^{(x)}$. After replacing into $\mathcal{L}$ the magnetization components $\mathcal{M}_x$, $\mathcal{M}_y$, and $\mathcal{M}_z$ (see Eq. 5.7) up to the second order, we obtain

$$\mathcal{L} \approx M_\mathrm{s} \frac{\partial m_Y}{\partial z} - M_\mathrm{s} \sin\phi_\mathrm{m} \frac{\partial m_X}{\partial x} - m_Y \cos\phi_\mathrm{m} \frac{\partial m_X}{\partial x} + m_X \cos\phi_\mathrm{m} \frac{\partial m_Y}{\partial x} \quad (5.22)$$

In the limit of a thin film, the DMI Hamiltonian becomes

$$\mathcal{H}_\mathrm{DM} = \frac{Dd\cos\phi_\mathrm{m}}{M_\mathrm{s}^2} \int \left[m_X \frac{\partial m_Y}{\partial x} - m_Y \frac{\partial m_X}{\partial x}\right] dx\, dz. \quad (5.23)$$

Within a Fourier representation of the dynamic magnetization, the frequency dispersion of the spin waves is separated into two contributions, $f(\mathbf{k}) = f_r(\mathbf{k}) + f_{DM}(\mathbf{k})$, a reciprocal part and a nonreciprocal one associated to the DMI, which is given by

$$f_{DM}(\mathbf{k}) = \frac{\gamma D(d_{HM})}{\pi M_s} k \sin\phi_{\mathbf{k}} \cos\phi_m. \tag{5.24}$$

In this expression $D(d_{HM}) \equiv S^2/(na_0 d)]|\mathbf{D}_{ij}|$ is the volume-averaged DMI strength, $\gamma$ is the gyromagnetic ratio, $a_0$ is the separation between nearest-neighbor ferromagnetic spins, and $n$ is an integer introduced to consider first, second, or even third neighbors [86]. The reciprocal part $f_r(\mathbf{k})$ includes exchange, dipolar, anisotropy, and Zeeman contributions [96]. The frequency difference of oppositely propagating spin waves, $\Delta f = f(\mathbf{k}) - f(-\mathbf{k})$, is then calculated as

$$\Delta f(\mathbf{k}, d_{HM}) = \frac{2\gamma D(d_{HM})}{\pi M_s} k \sin\phi_{\mathbf{k}} \cos\phi_m. \tag{5.25}$$

Most of the parameters involved in the frequency difference can be controlled by the experimental setup. For example, $\phi_m$ is controlled with the orientation of the applied field, $\phi_{\mathbf{k}}$ and $k$ are controlled with the excitation mechanism, while $\gamma$ and $M_s$ can be measured by conventional magnetometry or ferromagnetic resonance (FMR). Then, by measuring $\Delta f$, the strength of the interfacial DMI can be directly measured, as was theoretically proposed in Ref. [96], and then measured by several groups, mainly using BLS [72–87], as summarized in Tables 5.2 and 5.3. It is worth to mention that the estimation of the DMI strength through measurement of the frequency difference between counterpropagating spin waves, in the field-polarized case, should be the most reliable method, since there are no assumptions about the possible magnetic texture. Moreover, there is no need to neglect or approximate the dipolar interactions, as in the case of DW-based methods. As can be seen from Tables 5.2 and 5.3, some groups have measured the DMI strength with both BLS and DW-based methods, where different values of $D$ are found [81, 127].

The theoretical model developed previously has been applied to analyze experimental observations of the strength of the DM constant in Pt($d_{HM}$)/CoFeB films. Specifically, Tacchi et al. [86] studied the influence of the heavy metal thickness ($d_{HM}$) on the induced

**Table 5.2** Reported magnitudes for DM couplings in various heavy metal/ferromagnet interfacial systems. Experimental measurements include BLS, electrical and domain wall (DW)-based methods, micromagnetic simulations, and microscopy techniques. The thickness of each layer comprising the structure is measured in nanometers (nm).

| Material | Reference | $|D|$ (mJ/m$^2$) | Method |
| --- | --- | --- | --- |
| Pt(3)/CoFe(0.6)/MgO | [37] | 0.5 | DW |
| Pt(5)/Co(0.7)/Ir(0.46)/Pt(3) | [118] | 0.12 | DW |
| Pt(4)/Co(1.6)/Ni | [72, 73] | 0.44 | BLS |
| Pt(2)/CoFeB(0.8)/MgO | [74] | 1.0 | BLS |
| Pt(4)/CoFeB(1–1.6)/AlO$_x$ | [75] | 0.8−0.4 | BLS |
| Pt(4)/Co(1–2)/AlO$_x$ | [75] | 1.2−0.9 | BLS |
| Pt(6)/NiFe(1.3–10)/SiN | [76] | 0.15−0.025 | BLS |
| Pt(6)/NiFe(4)/Si | [77] | 1.2 | BLS |
| Pt(3)/Co(0.6–1.2)/AlO$_x$ | [78] | 2.7–1.57 | BLS |
| Ta/Pt(4)/Co(1.35)/AlO$_x$ | [79] | 1.65 | BLS |
| Pd/Fe/Ir | [119] | 3.9 | SP-STM/DW Simulation |
| (Ir(1)/Co(0.6)/Pt(1))$_{10}$ | [41] | 1.6−1.9 | STXM/DW Simulation |
| Ir(4)/Co(1.2–3)/AlO$_x$ | [80] | 0.3−0.7 | BLS |
| W(2-3)/CoFeB(1)/MgO | [81] | 0.25−0.27 | BLS |
| W(2-3)/CoFeB(1)/MgO | [81] | 0.23−0.12 | DW |
| TaN(1)/CoFeB(1)/MgO | [81] | 0.31 | BLS |
| TaN(1)/CoFeB(1)/MgO | [81] | 0.05 | DW |
| Hf(1)/CoFeB(1)/MgO | [81] | 0.15 | BLS |
| Hf(1)/CoFeB(1)/MgO | [81] | 0.01 | DW |
| Pt(3)/Co(16)/MgO | [120] | 0.25 | Electrical |
| Ta(5)/CoFeB(0.8)/MgO(2) | [121] | 0.057 | Kerr/DW |

DMI by means of BLS measurements of ultrathin CoFeB films. As a result, it was found that the strength of the interfacial DMI increases with heavy metal thickness, reaching a saturation value for $d_{HM}$ larger than a few nanometers. In the BLS measurements reported in Ref. [86], the Stokes and anti-Stokes peaks, corresponding to spin waves propagating in opposite directions, are simultaneously observed with comparable intensity. These peaks are characterized by a frequency shift that increases with heavy metal thickness. The frequency of both peaks interchanges when reversing the direction of the applied magnetic field, due to the reversal of the propagation direction of the spin waves. It was also found that the frequency asymmetry exhibits a linear dependence as a function of *k*, and it becomes more pronounced when increasing the thickness of the heavy metal. The in-plane angular dependence of the frequency shift was also measured with BLS [73, 75, 86, 87], which evidences a clear sinusoidal dependence with $\phi_\mathbf{k}$, in agreement with Eq. 5.25, as predicted theoretically [96].

**Table 5.3** Continuation of Table 5.2. The magnitudes reported in the section at the bottom of this table refer to DMI values from atomistic parameters or spatial units.

| Material | References | $|D|$ (mJ/m$^2$) | Method |
|---|---|---|---|
| Pt(3)/CoFe(1–1.6)/MgO | [82] | 1.4–2.3 | BLS |
| Pt(4)/CoFe(0.8–1.6)/MgO | [82] | 0.9–1.5 | BLS |
| W/CoFeB(1)/SiO$_2$ | [83] | 0.25 | BLS |
| Ta(5)/CoFeB(1)/Pt(0.15)/MgO | [84] | 0.056 | BLS |
| Pt(0.4–6)/CoFeB(2)/Cu | [86] | 0–0.45 | BLS |
| Co(1.4)/Pt | [122] | >0.5 | DW |
| Pt(5.4)/Co(2.5)/Au,Ir(0–2.5)/Pt(2.6) | [123] | 0.6 | BLS |
| Ta/CoFeB/TaO$_x$ | [87] | 0.22 | BLS |
| Ir/Co$_2$FeAl(0.9)/Si | [124] | 0.34 | BLS |
| MgO(5)/Fe(3)/Pt(4) | [125] | 0.349 | BLS |
| SiO$_2$/Fe(3)/Pt(4) | [125] | 0.225 | BLS |
| Pt(6)/MgO(367)/Fe(3)/Pt(4) | [125] | 0.335 | BLS |
| Pt(6)/MgO(367)/Fe(3)/Al(5)/SiO$_2$(5) | [125] | 0.1025 | BLS |
| Pt(6)/Co(3) | [126] | 0.33–0.43 | BLS |
| Pt/Co(0.8)/Ir(0–2)/Ta | [127] | 0.75–1.68 | BLS |
| Pt/Co(0.8)/Ir(0–2)/Ta | [127] | 0.48–1.25 | DW |
| Graphene/Ni$_{80}$Fe$_{20}$(3)/Ta(2) | [128] | 0.067 | BLS |
| FeFe/W | [129] | 7.5 meV/nm | SP-STM simulation |
| FeFe/W | [70] | 0.5–0.9 meV | SPEELS |
| (Co/Ni)$_n$/Ir/Pt | [35] | 0.12 meV/atom | BLS |
| (Ni/Co)$_n$/Ir | [130] | 0.36 meV/atom | BLS |
| Graphene/Co/Ru | [131] | 0.11 meV/atom | SPLEEM/DW |

## 5.4 Micromagnetic Simulations of Spin Waves with Interfacial DMI

Micromagnetic simulations are a powerful numerical method for the analysis of spin dynamics in magnetic samples of a few microns. In particular, this numerical technique can be used to calculate the spectrum of spin waves, and when a DMI is present, it reveals different properties characteristic of these systems, such as the spin-wave asymmetry. In this context, simulations have successfully supported theoretical formulations [80, 96, 97, 99, 132–137] and experimental measurements [72–87, 138] on thin-film systems with a homogeneous DMI.

A micromagnetic simulation is based on numerically discretizing the continuum description of the magnetic system into a mesh of magnetic moments whose arrangement depends on the discretization method. The two most commonly used methods are finite differences and finite elements. For the former, the sample is divided into a regular grid of cuboids, which each represents a magnetic moment (see Fig. 5.2a), and the expressions of the magnetic interactions are approximated according to this numerical method. The spacing between cuboids is usually chosen with a magnitude smaller than the exchange length. To simulate the dynamics of the magnetization, the Landau–Lifshitz–Gilbert equation of motion is numerically integrated for every magnetic moment. Three publicly available finite difference micromagnetic simulation codes are OOMMF [139], MuMax3 [140], and Fidimag [141].

The method for the simulation of spin waves propagating in a single direction has been discussed in detail in Refs. [142, 143] and in Ref. [144] for the case of a thin film with DMI. Following the techniques of the aforementioned studies, a similar system has been simulated (see Fig. 5.2) using the parameters shown in Table 5.4. To simulate the infinite film, it is customary to specify a sufficiently

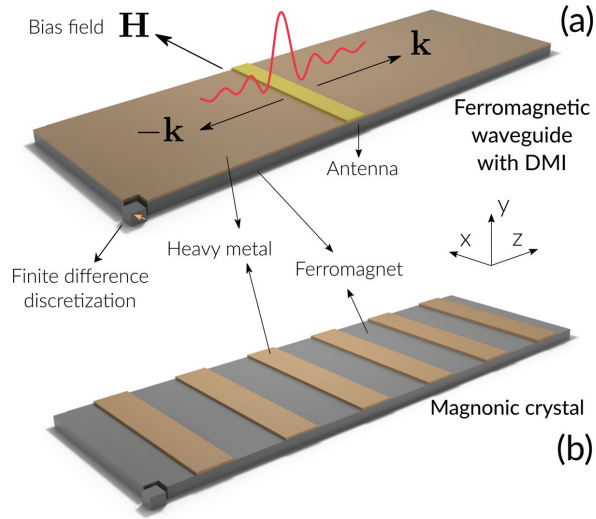

**Figure 5.2** (a) Thin-film waveguide system with interfacial DMI, where a ferromagnetic film is in contact with a heavy metal surface and the sample is saturated with a bias field. An antenna at the center of the sample can be used to excite counterpropagating spin waves (characterized by a spin-wave vector **k**) across the waveguide. Micromagnetic simulations are performed using a regular grid of cuboids representing magnetic moments. (b) Magnonic crystal specified by a regular pattern of heavy metal stripes.

long waveguide along the spin-wave propagation direction, in this case the *x* direction. Consequently, the sample is saturated with a bias field **H**. To excite the spin waves, a weak periodic magnetic field $\mathbf{h}_{\text{exc}}$ is applied in a region at the center of the waveguide in order to generate waves propagating in the two opposite directions, and it is applied perpendicular to the bias field. Experimentally, the spin-wave pulses can be generated using an antenna [145] (see Fig. 5.2a). The amplitude of the simulated periodic field is significantly smaller than the bias field. It is suggested to use a sinc function dependent on time for the field, delayed by time $t_0$, in order to excite all possible wave modes within a specific range of frequencies given by a cut-off frequency $f_0$ in the sinc function argument. A similar method can be applied to the spatial coordinates to restrict the range of wave vector magnitudes [142]. When exciting the system, the magnetization components are saved for every magnetic moment in regularly spaced time steps $\Delta \tau$. Finally, after a sufficiently long time $\tau$, a 2D Fourier transform in space and time variables is performed in order to obtain the spin-wave spectrum. To avoid noise in the spectrum caused by the reflection of the spin waves at the sample edges, it has been shown that using a strong damping towards the sample edges minimizes this effect [146]. Alternatively, it is also possible to use periodic boundary conditions.

**Table 5.4** Parameters for the simulation of spin waves propagating in a thin-film waveguide with DMI and with periodicity along the $y$ direction. Dimensions are $2000 \times 128 \times 1$ nm$^3$.

| Symbol | Magnetic parameter | Value |
| --- | --- | --- |
| $A$ | Exchange | 11.1 pJ/m |
| $D$ | Dzyaloshinskii–Moriya strength | 3.0 mJ/m$^2$ |
| $M_s$ | Saturation magnetization | 658 kA/m |
| $\mu_0 \mathbf{H}$ | External field | (0, 0.25, 0) T |
| $\mu_0 \mathbf{h}_{exc}$ | Excitation field | (0.025, 0, 0) T |
| $f_0$ | Cut-off frequency | 60 GHz |
| $t_0$ | Excitation delay | 50 ps |
| $\gamma$ | Gyromagnetic ratio | $1.76 \times 10^{11}$ Hz/T |
| $\alpha$ | Gilbert damping parameter | 0.01 |
| $\tau$ | Excitation time | 4 ns |
| $\Delta \tau$ | Saving time step | 0.5 ps |

The result of the simulated spin-wave spectrum in a thin film of permalloy, specified in Table 5.4, is depicted in Fig. 5.3, where the spin waves propagate perpendicular to the saturation field. The excited modes in this case are known as DE modes. This simulation was obtained with the MuMax3 code using periodic boundary conditions along the waveguide width. The spin-wave spectrum is shown as the intensity of the 2D Fourier transform of the $x$ component of the magnetization, which is scaled logarithmically, and is directly compared with the theory [96, 97]. A test for the accuracy of the simulation is the estimation of the DMI from the spin-wave asymmetry. Accordingly, a second-order polynomial is fitted using the data from $k = -0.0785$ nm$^{-1}$ up to $k = 0.2985$ nm$^{-1}$, and the asymmetry predicts a DMI magnitude of 2.89 mJ/m$^2$, in good agreement with the original value of 3 mJ/m$^2$. It is possible to extend the lower limit of the data for the curve fit; however, the polynomial

**18** *Spin Waves in Thin Films and Magnonic Crystals with Dzyaloshinskii–Moriya Interactions*

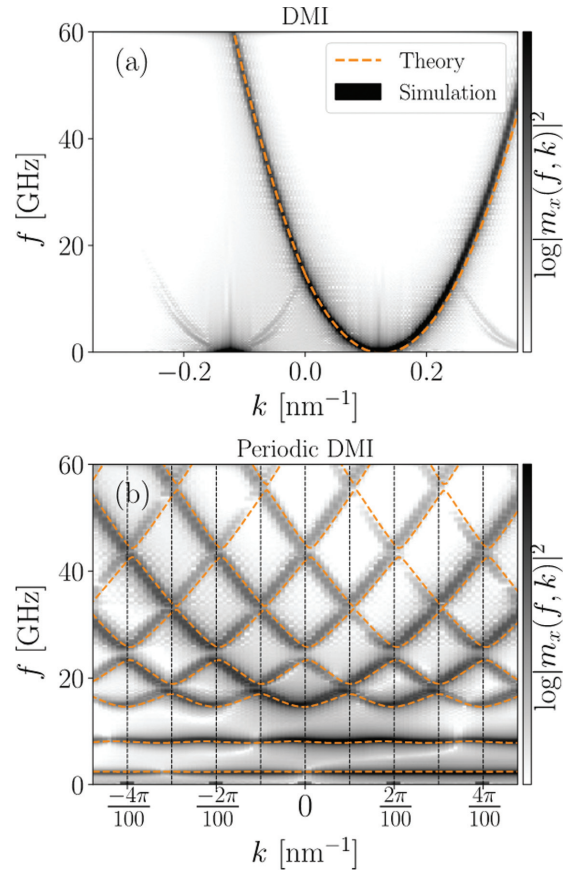

**Figure 5.3** (a) Micromagnetic simulations of spin-wave propagation in permalloy thin films with (a) homogeneous DMI from a heavy metal film on top of the ferromagnet and (b) periodic DMI from regularly patterned heavy metal stripes. Simulations were obtained with the MuMax3 code. Details of simulation (a) are specified in Table 5.4. Simulation (b) is similar, but the DMI is set in regions of 50 nm width with a periodicity of 100 nm.

becomes less accurate in reproducing the intensity peaks away from the minimum, which can be improved by setting a larger simulation time. Moreover, since a cut-off frequency was set to 60 GHz, the spectrum is approximately valid down to $k \approx -0.1$ nm$^{-1}$.

From the result discussed previously, it can be concluded that micromagnetic simulations are highly accurate to reproduce the theoretical predictions. It is now straightforward to extend the simulations to more complex scenarios. For instance, in Fig. 5.3b the same permalloy system is simulated by setting a periodic DMI along the long axis of the waveguide. This can be obtained by patterning the top or bottom surface of the ferromagnetic system with regularly spaced heavy metal wires [147]. The simulated system of Fig. 5.3b was specified by alternating regions of material with and without a DMI of 50 nm width. In this case, the periodic DMI induces flat bands and indirect bandgaps, which are not characteristic of a magnonic system [147]. These effects are thoroughly discussed in Section 5.5.

## 5.5 Spin Waves in Thin Films with a Periodic DMI

Spin waves propagating in a periodic magnetic media, such as an MC [148]—the magnetic counterpart of a photonic crystal—exhibit magnonic bandgaps induced by Bragg scattering, which means there are frequencies where propagation is forbidden. These bandgaps appear at the boundaries of the Brillouin zones (BZs) and are induced when waves with wave vector $\mathbf{k} = \pm n\pi/a$, with $a$ being the lattice parameter and $n$ an integer that satisfies the Bragg condition. On the other hand, the Bragg condition is modified by the breaking of symmetry induced by DMIs, since two counterpropagating waves at the same frequency have different wavelengths. Therefore, neither magnonic bandgaps nor the standing waves appear at the BZ edges anymore. According to this, MCs with a periodic chiral interaction exhibit indirect bandgaps, where band edges lie outside the borders of the BZs. Different cases of MCs with chiral properties have been studied recently [100, 101], reporting the observation of indirect bandgaps. Topological properties were also found in a theoretical study of periodic arrays of magnetic nanoislands on top of a heavy metal layer [149]. In addition, it has been shown, by means of micromagnetic simulations, how the DMI induces a nontrivial temporal evolution of standing spin waves in a geometrically confined structure [150]. Recently, it has been shown how a periodic DMI can produce all these effects in a magnonic thin-film system [147], namely (i) indirect bandgaps, (ii) low-frequency flat bands, and (iii) an unconventional temporal evolution of the standing waves at the bandgap edges. Following this latter work, this section is dedicated to the theoretical background of the role of a periodic DMI.

An overview of a chiral MC is shown in Fig. 5.2b, where a periodic array of heavy metal stripes is put in contact with a ferromagnetic layer. Owing to the high spin–orbit coupling of the heavy metal and the ultrathin thickness of the ferromagnetic layer, a periodic DMI is expected to occur in the system. In the continuum limit, effective fields associated with the different interactions are required to study the dynamic of the MC by means of both the Landau–Lifshitz equation and the plane-wave method. In the latter approach, the periodic quantities are expanded into Fourier series so that the equation of motion is treated as an eigenvalue problem [147]. While the effective fields associated with anisotropies, dipolar interactions, and exchange interactions are very well known, the effective field due to DMI ($\mathbf{h}^{\text{DM}}$) becomes nontrivial, because the DMI strength ($D$) is now a space-dependent periodic parameter. To derive the DM effective field, it is necessary to start from the atomic Hamiltonian $\mathcal{H}_{\text{DM}} = \sum_{\langle i,j \rangle} \mathbf{D}_{ij} \cdot (\mathbf{S}_i \times \mathbf{S}_j)$. Since the DMI is short ranged, one can see that only the terms $\mathbf{D}_{i-1,i} \cdot (\mathbf{S}_{i-1} \times \mathbf{S}_i) + \mathbf{D}_{i,i+1} \cdot (\mathbf{S}_i \times \mathbf{S}_{i+1}) = \mathbf{S}_i \cdot (\mathbf{D}_{i-1,i} \times \mathbf{S}_{i-1} - \mathbf{D}_{i,i+1} \times \mathbf{S}_{i+1})$ are involved in $\mathcal{H}_{\text{DM}}$, thus

$$\mathcal{H}_{\text{DM}} = \sum_i \mathbf{S}_i \cdot (\mathbf{D}_{i-1,i} \times \mathbf{S}_{i-1} - \mathbf{D}_{i,i+1} \times \mathbf{S}_{i+1}). \quad (5.26)$$

Here, $\mathbf{S}_{i\pm 1}$ represent the neighbor spins of $\mathbf{S}_i$, and the following effective field for the spin at site *i* is obtained:

$$\mathbf{h}_i^{\text{DM}} = \mathbf{D}_{i-1,i} \times \mathbf{S}_{i-1} - \mathbf{D}_{i,i+1} \times \mathbf{S}_{i+1}. \quad (5.27)$$

In a very extended array of spins, the variation of the spin orientation between neighboring sites is very smooth; therefore, in the continuum approximation, the elements located around site *i* can be written as

$$\mathbf{S}_{i\pm 1} \simeq \mathbf{S}_i \pm \frac{\partial \mathbf{S}_i}{\partial z} \delta z, \quad (5.28)$$

$$\mathbf{D}_{i,i+1} \simeq \mathbf{D}_{i-1,i} + \frac{\partial \mathbf{D}_{i-1,i}}{\partial z} \delta z. \quad (5.29)$$

Notice that it has been assumed that the DM coupling only varies along the *z* axis (see Fig. 5.2b); thus, the effective field in Eq. 5.27

becomes

$$\mathbf{h}_i^{\text{DM}} \simeq \left( -2\mathbf{D}_{i-1,i} \times \frac{\partial \mathbf{S}_i}{\partial z} - \frac{\partial \mathbf{D}_{i-1,i}}{\partial z} \times \mathbf{S}_i \right) \delta z. \quad (5.30)$$

Taking into account that $\mathbf{D}_{i-1,i} \perp (\mathbf{r}_i - \mathbf{r}_{i\pm1})$, the effective DMI field in the continuum limit is

$$\mathbf{h}^{\text{DM}} \simeq \frac{1}{\mu_0 M_s^2} \left\{ \left[ 2D(\mathbf{r}) \frac{d}{dz} m_z(\mathbf{r}) + m_z(\mathbf{r}) \frac{d}{dz} D(\mathbf{r}) \right] \hat{y} \right.$$
$$\left. - \left[ 2D(\mathbf{r}) \frac{d}{dz} m_y(\mathbf{r}) + m_y(\mathbf{r}) \frac{d}{dz} D(\mathbf{r}) \right] \hat{z} \right\}, \quad (5.31)$$

where $D(\mathbf{r})$ is the DM strength that is a periodic function of $z$ and can be expanded as $D(\mathbf{r}) = \sum_{\mathbf{G}} D(\mathbf{G}) e^{i \mathbf{G} \cdot \mathbf{r}}$, with $\mathbf{G} = (2\pi n/a)\hat{z}$ a reciprocal lattice vector. By including all the energy contributions into the Landau–Lifshitz equation [151], the following eigenvalue problem is obtained:

$$\tilde{\mathbf{T}} \, \mathbf{m}_{\mathbf{G}} = i \frac{\omega}{\mu_0 \gamma} \mathbf{m}_{\mathbf{G}}, \quad (5.32)$$

where $\mathbf{m}_{\mathbf{G}}^{\text{T}} = [m_z(G_1), \ldots, m_z(G_N), m_y(G_1), \ldots, m_y(G_N)]$ is the eigenvector and $\tilde{\mathbf{T}}$ is given by

$$\tilde{\mathbf{T}} = \begin{pmatrix} \tilde{\mathbf{T}}^{zz} & \tilde{\mathbf{T}}^{zy} \\ \tilde{\mathbf{T}}^{yz} & \tilde{\mathbf{T}}^{yy} \end{pmatrix}. \quad (5.33)$$

The matrix elements of the submatrices are defined as follows:

$$\mathbf{T}_{\mathbf{G},\mathbf{G}'}^{zz} = \mathbf{T}_{\mathbf{G},\mathbf{G}'}^{yy} = -i \frac{2D(\mathbf{G} - \mathbf{G}')}{\mu_0 M_s} \left( \frac{\mathbf{G} + \mathbf{G}'}{2} + \mathbf{k} \right) \cdot \hat{z}, \quad (5.34a)$$

$$\mathbf{T}_{\mathbf{G},\mathbf{G}'}^{zy} = -\left[ M_s (\mathbf{G} + \mathbf{k})^2 \lambda_{\text{ex}}^2 + H + M_s F_{\mathbf{G},\mathbf{k}} \right] \delta_{\mathbf{G},\mathbf{G}'}, \quad (5.34b)$$

$$\mathbf{T}_{\mathbf{G},\mathbf{G}'}^{yz} = \left[ M_s (\mathbf{G} + \mathbf{k})^2 \lambda_{\text{ex}}^2 + H + M_s (1 - F_{\mathbf{G},\mathbf{k}}) \right] \delta_{\mathbf{G},\mathbf{G}'}. \quad (5.34c)$$

In Eq. 5.34, $F_{\mathbf{G},\mathbf{k}} = e^{-|\mathbf{G}+\mathbf{k}|d/2}$, $\lambda_{\text{ex}}$ is the exchange length and $d$ is the thickness of the ferromagnet.

To understand the dynamic properties of the system shown in Fig. 5.2b, where the equilibrium magnetization is pointing along $x$ and the spin wave propagation is along $z$, typical magnetic parameters of permalloy are used (see Table 5.4). In the zero-DMI case, the dispersion is completely reciprocal, as shown in Fig. 5.4a.

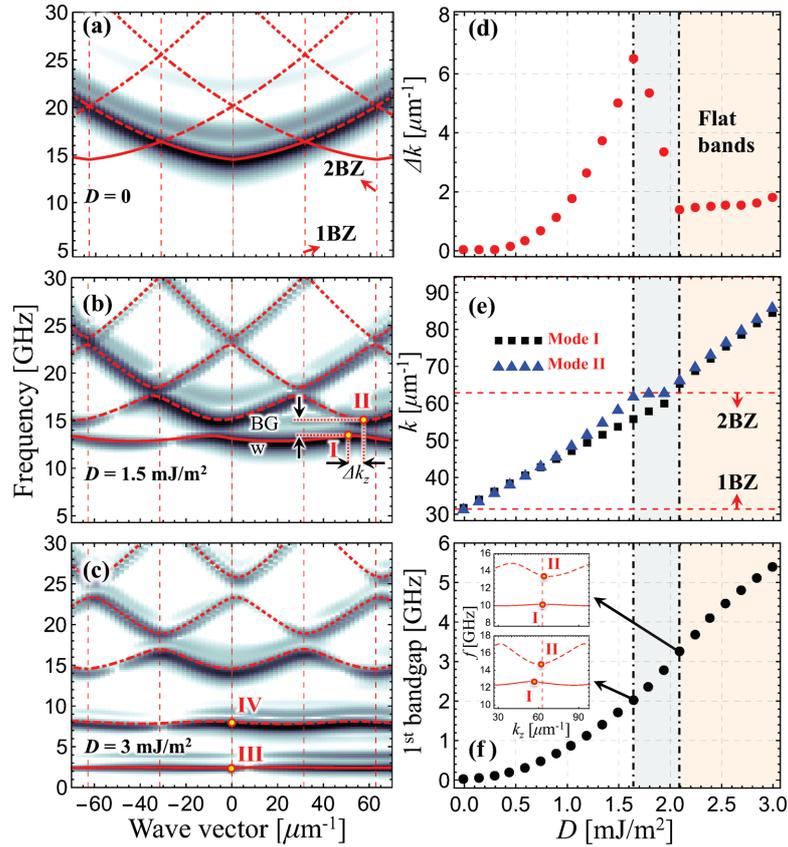

**Figure 5.4** (a–c) Dispersion relation for a Py film covered with an array of heavy-metal wires, where $a = 100$ nm, $D = 0$, 15 and 3 mJ/m$^2$. The lines correspond to the theory and the color code to simulations with the OOMMF code, where darker (lighter) color represents an intensity maximum (minimum). Panel (d) depicts the wave-vector shift $\Delta \mathbf{k}$, while (e) illustrates the modes I and II [see (b)] as a function of $D$. In (f) the first bandgap as a function of the Dzyaloshinskii–Moriya strength $D$ is shown. The shaded areas correspond to three ranges of $D$-values where different behaviors are predicted. Insets in (f) show the first bandgap behavior during the transition between the shaded areas [147].

When a periodic DMI is present, indirect bandgaps emerge, which are generated by the nonreciprocal character of the DMI. Interestingly, when the DMI strength is increased the low-frequency branches decrease in frequency and become flat (see Fig. 5.4c,d). One particular feature of the spin-wave dynamics with a periodic DMI is that the transition from reciprocal dispersion (see Fig. 5.4a) up to the emergence of low-frequency flat bands is not monotonic, as illustrated in Fig. 5.4d,f. When the parameter $D$ is increased, the indirect character of the spin waves also increases. Then, the mode II (minimum of the bandgap edge shown in Fig. 5.4b) reaches the second Brillouin zone (2BZ), and it remains pinned under the increasing of $D$, as shown by the triangles in Fig. 5.4e. At this point, mode I is still shifting to high $k$ values, so when mode I also reaches the 2BZ, the flat region is reached and the mode shifts to lower frequencies. If $D$ further increases, the upper-frequency branches also become flat. The emergence of these flat bands is enhanced if the lattice parameter increases (not shown), and it can be manipulated by changing the geometrical properties of the heavy metal stripes [147]. These flat bands possibly have important consequences, such as the Bose–Einstein condensation [152] predicted in bosonic systems hosting flat bands [153], and frustrated quantum magnets [154, 155]. Topological properties of higher-dimensional periodic DMI structures are still to be explored.

It is important to remark that the influence of a continuous heavy metal film in contact with a ferromagnetic layer is not observed in the FMR response [96]. The inclusion of a periodic DMI modifies the band structure in a way that additional modes can be now observed in FMR experiments [101, 147, 150]. A difficulty arises, however, when trying to understand the temporal evolution of these waves, because the standing waves around the gaps originate outside the BZs. Indeed, while a typical standing behavior is observed in the zones without DMI, the spin waves have a nonzero phase velocity in the areas in contact with the heavy metal stripes. This time dependence can be understood by taking into account the boundary conditions induced by DMI at the interfaces, since the DMI produces canted states of the dynamic magnetization, inducing, thus, a nontrivial temporal evolution of the spin waves along the whole system [147]. Furthermore it can be shown that the inclusion of Gilbert damping does not significantly modify the band structure obtained in Fig. 5.4. However, it is well known that if a ferromagnet is coupled to a heavy metal, the damping increases, decreasing then the propagation length of the waves, which is not favorable for applications. Hence, other kinds of nonreciprocal sources could be of interest, as, for instance, (i) an electric current using the adiabatic spin transfer torque [156, 157]; (ii) the flexoelectric interaction [158]; (iii) metallized MCs [159, 160]; (iv) a bicomponent MC composed of a normal ferromagnet and a bulk periodic DMI material; (v) periodic ferromagnetic bilayers, where nonreciprocity is induced by dynamic dipolar interaction [62, 68, 161–165]; or even (vi) films with graded magnetization along the thickness [166].

## 5.6 Conclusions

The influence of DMI on the behavior of spin waves in chiral lattice magnets, ultrathin ferromagnetic films capped with a heavy metal layer, and chiral MCs has been reviewed. Such anisotropic exchange coupling produces nonreciprocal features on the spin-wave spectrum of a magnetic system, a phenomenon that occurs for both bulk and interfacial DMI. For bulk DMI systems, such as MnSi, FeGe, and $Cu_2OSeO_3$, it was theoretically shown that spin-wave nonreciprocity should appear mostly in backward-volume configuration. On the other hand, the SW nonreciprocity induced in thin films with interfacial DMI, as, for instance, Pt/Co/Ni or Ir/Co/AlO$_x$, is maximum in the DE geometry. Interestingly, for both types of DMI, the frequency nonreciprocity is proportional to the wave vector, the strength of the DMI, and the in-plane component of the static magnetization, vanishing for perpendicularly magnetized thin films. In the case of interfacial DMI, it has been also shown that the frequency shift decreases with the thickness of the ferromagnetic film, as a consequence of the interfacial nature of the coupling, while the nonreciprocity increases with the thickness of the heavy metal layer and saturates in a length scale of a few nanometers. This behavior is a result of the spin–orbit coupling between the sea of conduction electrons and the heavy metal atoms around the ferromagnetic sites.

The concept of a chiral MC has been introduced theoretically, where the interfacial DMI is considered as periodic along a single direction. The effect of this periodicity includes additional features in the magnon spectrum such as flat bands, indirect gaps, and an unusual spin-wave evolution, with standing waves showing finite phase velocities in the zones where the DMI is nonzero. These results have been obtained with micromagnetic simulations and using a theoretical approach based on the plane-wave method. These chiral MCs with periodic DMI, which may be attained, for instance, by covering a thin ferromagnetic film with an array of heavy metal wires, host interesting physical properties, encouraging future experimental studies to prove and evidence these phenomena.

## Acknowledgments

We acknowledge fruitful discussions with R. Arias, D. L. Mills, T. Schneider, J. Lindner, K. Lenz, A. Roldán-Molina, G. Gubbiotti, S. Tacchi, M. Ahlberg, H. Fangohr, and F. Ma. This work was supported by FONDECYT grants 1161403, 11170736, and 3150372 and Centers of Excellence with Basal/CONICYT financing, grant FB0807, CEDENNA. D. C.-O. acknowledges support from the EPSRC Programme grant on Skyrmionics (EP/N032128/1).

May 28, 2019 12:35 JSP Book - 9in x 6in 05-Gubbiotti-05